\documentstyle [12pt]{article}
\begin{document}

\hfill {WM-94-107}

\hfill {June, 1994}

\vskip 1in   \baselineskip 24pt
{
\Large
   \bigskip
   \centerline{SUSY GUTs without light
Higgs bosons} }
 \vskip .8in

\centerline{Marc Sher}  \bigskip
\centerline {\it Physics Department, College of William and Mary}
\centerline {\it Williamsburg, VA 23187, USA}
\vskip 1in

{\narrower\narrower There are two different types of perturbative
unification in SUSY GUTs.  Perturbative gauge unification is
necessary for the extremely successful prediction of
$\sin^2\theta_w$.  ``Perturbative validity'' requires that {\it
all} couplings (not just the gauge couplings) remain small; this
assumption is essential to the proof that any SUSY GUT must have
a light Higgs with a mass below $140$ GeV.ÊÊ It is pointed out
here that if the latter bound is violated (so that there need be
no light Higgs bosons), then it is plausible that the successful
prediction of $\sin^2\theta_w$ will {\it not} be significantly
altered.

}

\newpage

\def\beq{\begin{equation}}
\def\eeq{\end{equation}}

\def\st{\sin^2\theta_w}

An impressive successful prediction of supersymmetric grand unified
 theories (SUSY GUTs) is the prediction of the value of $\st$.  If
one assumes that, as expected, the  scale of supersymmetry breaking
is within an order of magnitude of the electroweak scale, then the
predicted value of $\st$ is \cite{langacker} $0.2340\pm 0.0026$,
compared with the measured value of
$0.2329\pm 0.0006$.  This has led to a resurgence of interest in
such theories.

Another prediction of SUSY GUTs concerns the mass of the lightest
Higgs boson.  It has been shown \cite{KKW} that an upper limit to
this mass exists in {\it any ``perturbatively valid''
supersymmetric grand unified theory}, independent of the gauge,
fermion and/or scalar structure of the theory, and the bound has a
value of approximately
$140$ GeV.  This gives such models the (almost) unique ability to be
experimentally excluded.

How does this general upper bound arise?  In a general
supersymmetric model, there will be a number of arbitrary coupling
constants in the superpotential.  Although the minimal
supersymmetric model has no such couplings, the simplest extension
(with a singlet $N$ added) has a $\lambda H\overline{H}N$ term in
the superpotential, where
$\lambda$ is arbitrary; more complicated models may have many such couplings.
The argument of Ref. [2] is based on the requirement that a model
is only ``perturbatively valid'' if {\it all} of the couplings
(including the various
$\lambda$'s) stay relatively small between the electroweak and
unification scales.  For example, in the simplest extension, the
requirement that $\lambda$ not blow up before the unification scale
is reached places an upper bound on $\lambda$ at the electroweak
scale, which leads directly to an upper bound on the mass of the
lightest Higgs boson.

In this Letter, I will consider the consequences of violating this
upper bound (significantly, not by just a few tens of GeV),
concentrating on the simplest extension of the minimal
supersymmetric model.  In this case, $\lambda$ will diverge before
the unification scale is reached. Of course,
$\lambda$ doesn't really diverge; rather, it becomes strong and the
new strongly interacting theory has new physics, the nature of which
is unknown.  I will argue that it is possible that this new physics
 will NOT appreciably affect the running of the gauge coupling
constants, and thus will not affect the successful prediction of
$\st$.  Thus, if the successful prediction of $\st$ is due to
perturbative unification of a supersymmetric model, this will not
automatically imply the existence of a light Higgs boson. In short,
I am distinguishing between perturbative gauge unification (which
gives the prediction of $\st$) and perturbative validity in all
couplings (which is needed for the upper bound on the lightest
Higgs boson).

Before looking at supersymmetric models, first consider the
familiar (non-supersymmetric) standard model, and suppose that the
minimal $SU(5)$ prediction of $\st$ had turned out to be
experimentally correct.  One can also obtain a bound of about $180$
GeV on the Higgs mass by requiring that the scalar self-coupling
not diverge by the unification scale.  In the standard model,
however, the scalar self-coupling only enters the beta-functions
for the gauge couplings at {\it three-loop} order.   As a result,
even if the self-coupling becomes very large, and new physics
enters, the effect on the gauge-coupling beta-functions will not be
much bigger than the two-loop contributions\footnote{Unless, as
will be noted later,  the new physics involves a composite model
in which the new states have $SU(3)\times SU(2)\times U(1)$ quantum
numbers, in which case these states will affect the beta functions
at one-loop.}, and will thus be very small.  Thus, in the minimal
standard model,  the prediction of
$\st$ will not be significantly affected if the scalar sector
becomes strong at some intermediate scale.  In supersymmetric
models, however, the additional parameters are Yukawa couplings,
and thus enter at two-loop order.

\def\sx{{1\over 16\pi^2}} To begin, consider the simplest
extension, with a singlet $N$ and a
$\lambda H\overline{H}N$ superpotential term.  We will ignore a
possible $kN^3$ term since that will tend to decrease the upper
bound to the lightest Higgs mass \cite{king}.  To leading order in
the gauge couplings, the renormalization group equations for the
gauge couplings are \cite{vaughn} \beq{dg_3\over dt}=\sx
b_3g^3_3\eeq\beq {dg_2\over dt}=\sx b_2g^3_2-{1\over
(16\pi^2)^2}g^3_2\lambda^2\eeq\beq {dg_1\over dt}=\sx
b_1g^3_1-{3\over 5}{1\over (16\pi^2)^2}g^3_1\lambda^2\eeq where
$b_3=-3;\ b_2=1;\ b_1=6.6$.  These equations are derived under the
assumption that $\lambda$ is in the perturbative region, of
course.  Suppose that $\lambda$ is not small, however.  Then the
right hand sides of the above equation can be replaced
\cite{hold,Holdom} by
$2\Pi(q^2)g^3$, where the gauge boson self-energy is written as $
\Pi_{\mu\nu}(q)=g^2\Pi(q^2)g_{\mu\nu}+q_\mu q_\nu {\rm terms}.$ We
could thus define
$\lambda$ in terms of the self-energy such that the above equations
are valid; and then keep in mind that it is equal to the coupling
constant in the superpotential {\it only} when $\lambda$ is small.

 The above equations can be separated, and the low energy value of
$\st$ can be determined.  The result (again, to leading order in
the {\it gauge} couplings) is
\beq\label{sintheta}\st={1\over 6}+{5\alpha\over
9\alpha_s}+\left({5\over 18}-{20\alpha
\over 27\alpha_s}\right)\left({2b_3-3b_2+b_1+{12\over
5}\sx\overline{\lambda}^2 \over {8\over 3}b_3-b_2-{5\over 3}b_1+2\sx
\overline{\lambda}^2}\right)
\eeq where
\beq\label{lam}\overline{\lambda}^2\equiv {\int_0^{t_X}\
\lambda^2(t)\ dt\over t_X},\eeq where $t_X\equiv \log(M_X/M_Z)$.
Thus, $\overline{\lambda}^2$ is the ``average'' value of $\lambda^2$
between the electroweak and unification scale.  Now, suppose we do
not wish the presence of $\lambda$ to alter the successful
prediction of $\st$ by more than two standard deviations, i.e. by
more than
$0.0052$.  This will occur if and only if
\beq\label{lambound}{\overline{\lambda}^2\over 16\pi^2}\ < \
0.207\eeq  Obviously, if
$\lambda$ reaches a Landau pole before the unification scale, then
$\overline{\lambda}$ is infinite and this condition is not
satisfied.

As stated earlier, $\lambda$ does not actually diverge; when it gets
large, new physics enters.  This situation is very similar to
looking at the evolution of the fine structure constant from the
weak scale to LOW energies, including gluon exchange in quark
loops.  There, a two loop term enters proportional to $e^3g_s^2$,
and $g_s$ appears to diverge at a scale of a few hundred MeV.  In
that case, we know that new physics enters, and the relevant states
become hadrons, rather than quarks.  Here, however, we do not know
the nature of the new physics caused by $\lambda$ getting large; so
how can one determine  whether or not Eq. [\ref{lambound}] is
satisfied?

Suppose we consider two alternatives for the new physics which
enters when $\lambda$ is large.  In the first model, there is an
ultraviolet fixed point in the beta-function at some $\lambda_o$.
In the second, the strong coupling is an indication that some of
the fields are composite.

In the first model, the coupling $\lambda$ increases until
$\lambda_o$ is reached, and then remains fixed. It is reasonable to
suppose that the value of $\lambda_o$ is at or near the unitarity
bound (which signifies a breakdown in perturbation theory). There
are two different unitarity bounds which will be of interest.  The
bound on the Yukawa coupling of a heavy quark \cite{marciano} is
$\sqrt{8\pi/3}$, whereas that for a heavy lepton is approximately a
factor of two larger\cite{chanowitz}.
 This latter bound corresponds to a value of
$\lambda_{max}\sim 6$.  An even stronger bound can be obtained by
noting that the superpotential term will lead to a quartic
$\lambda^2H^2\overline{H}^2$ interaction.  The bound on a scalar
self-interaction is given\cite{marciano} by $8\pi/5$; corresponding
to a value of $\lambda_{max}=\sqrt{8\pi/5}\sim 2.5$.

The $\lambda$ parameter in this case in not quite the same
as a Yukawa coupling or a scalar self-coupling.
One expects this bound to be stronger due to the additional
channels available in a supersymmetric model.  Fortunately, the
unitarity
bound on this parameter in a particular $SU(5)\times U(1)$ model
{\it has} been calculated\cite{lopez}, and one can easily
extract the bound in this model from that work.  It is found
that, independent of any parameters, the bound is always less
than $2.5$, and can occasionally be significantly less than $2.5$.
The reader should note that this bound is a factor of two stronger
than cited in that work, due to a later and more precise treatment
of partial wave unitarity\cite{marciano,lopeztwo}.

Since $\overline{\lambda}^2$
must be less than $\lambda_o^2$, it is clear that Eq.
[\ref{lambound}] will be satisfied, and that  there will  not be a
significant effect on
$\st$.  The upper bound to the mass of the lightest Higgs
boson\cite{king}, $\lambda(M_Z)v$, can then be as large as several
hundred GeV.  Thus, in this model, the striking prediction of
$\st$, while it does imply perturbative unification of the gauge
couplings, does NOT imply perturbation theory is valid up to the
unification scale for all of the other couplings.

In the second model, one or more of the fields is composite. Since
it is possible that the constituents carry $SU(3)\times SU(2)\times
U(1)$ quantum numbers (as do techniquarks, for example), then at
scales above the compositeness scale these constituents will
contribute directly, at one-loop, to the beta functions.  One will
then be able to radically change the prediction of $\st$.  Here, the
recent work of  Holdom\cite{Holdom} in integrating gauge coupling
beta functions through both the QCD and technicolor thresholds
would prove useful.

Thus, we have two models for the new physics at the scale at which
$\lambda$ becomes large.  In one, there is no significant effect on
the prediction of $\st$; in the other, there could be a very large
effect.  Since we have no way of knowing, at present, which model
would be a more accurate description of Nature, we can only conclude
that the successful prediction of $\st$ {\it might} not be altered
significantly if the lightest Higgs boson is considerably heavier
than $140$ GeV.  SUSY GUTs do not necessarily imply a light Higgs
boson.

I thank Franz Gross for pointing out the importance of the
unitarity bound, Michael Peskin for a very useful conversation and
Jose Goity for comments on the manuscript.  This work was
supported by the National Science Foundation.

\def\prd#1#2#3{{\it Phys. ~Rev. ~}{\bf D#1} (19#2) #3}
\def\plb#1#2#3{{\it Phys. ~Lett. ~}{\bf B#1} (19#2) #3}
\def\npb#1#2#3{{\it Nucl. ~Phys. ~}{\bf B#1} (19#2) #3}
\def\prl#1#2#3{{\it Phys. ~Rev. ~Lett. ~}{\bf #1} (19#2) #3}

\bibliographystyle{unsrt}

\end{document}